\documentclass[12pt]{article}
\newcommand{\nc}{\newcommand}
\nc{\bib}{\bibitem}
\nc{\al}{\alpha}
\nc{\g}{\gamma}
\nc{\G}{\Gamma}
\nc{\D}{\Delta}
\nc{\eps}{\epsilon}
\nc{\la}{\lambda}
\nc{\La}{\Lambda}
\nc{\var}{\varphi}
\nc{\cg}{{\cal G}}
\nc{\pa}{\partial}
\nc{\nn}{\nonumber \\ }
\nc{\hf}{\frac{1}{2}}  
\nc{\dz}{\frac{dz}{2\pi i}}
\nc{\bin}[2]{\left (\begin{array}{c} {#1}\\ {#2} \end{array}\right )}
\nc{\ben}{\begin{equation}}
\nc{\een}{\end{equation}}
\nc{\bea}{\begin{eqnarray}}
\nc{\eea}{\end{eqnarray}}
\nc{\bra}[1]{\langle {#1}|}
\nc{\ket}[1]{|{#1}\rangle}
\newcommand{\Z}{\mbox{$Z\hspace{-2mm}Z$}}
\nc{\C}{\mbox{\hspace{1.24mm}\rule{0.2mm}{2.5mm}\hspace{-2.7mm} C}}
\nc{\Nat}{\mbox{\hspace{.04mm}\rule{0.2mm}{2.8mm}\hspace{-1.5mm} N}}

\nc{\HH}{\mbox{\hspace{.04mm}\rule{0.2mm}{2.8mm}\hspace{-1.5mm} H}}

\def\vvdots{\mathinner{\mkern1mu\raise1pt\vbox{\kern7pt\hbox{.}}\mkern2mu
 \raise4pt\hbox{.}\mkern2mu\raise7pt\hbox{.}\mkern1mu}}

\begin{document}

\topmargin -5mm
\oddsidemargin 5mm

\begin{titlepage}
\setcounter{page}{0}

\vspace{8mm}
\begin{center}
{\huge On $SU(2)$ Wess-Zumino-Witten models}\\[.4cm]
{\huge and stochastic evolutions}

\vspace{15mm}
{\Large J{\o}rgen Rasmussen}\\[.3cm] 
{\em Centre de recherches math\'ematiques, Universit\'e de Montr\'eal}\\ 
{\em Case postale 6128, 
succursale centre-ville, Montr\'eal, Qc, Canada H3C 3J7}\\[.3cm]
rasmusse@crm.umontreal.ca

\end{center}

\vspace{10mm}
\centerline{{\bf{Abstract}}}
\vskip.4cm
\noindent
It is discussed how stochastic evolutions may be connected to
$SU(2)$ Wess-Zumino-Witten models. 
Transformations of primary fields are generated by
the Virasoro group and
an affine extension of the Lie group $SU(2)$. The transformations
may be treated and linked separately to stochastic evolutions.
A combination allows one to
associate a set of stochastic evolutions to
the affine Sugawara construction. The singular-vector decoupling generating the 
Knizhnik-Zamolodchikov equations may thus be related to
stochastic evolutions. The latter are based on an infinite-dimensional
Brownian motion.
\\[.5cm]
{\bf Keywords:} Conformal field theory, stochastic evolutions,
  Wess-Zumino-Witten models, Knizhnik-Zamolodchikov equations.
\end{titlepage}
\newpage
\renewcommand{\thefootnote}{\arabic{footnote}}
\setcounter{footnote}{0}

\section{Introduction}

There is a tradition in the physics community 
for describing a broad class of two-dimensional critical systems
in terms of conformal field theory (CFT).
Schramm has introduced the celebrated
stochastic L\"owner evolutions (SLEs) \cite{Sch} as a mathematical
rigorous way of handling some of these two-dimensional
systems at criticality. 
The method involves the study of stochastic evolutions of conformal maps,
and has been developed further in \cite{LSW,RS}.
Recent reviews on SLE may be found in \cite{Law,KN}.
Applications as well as formal properties and generalizations of SLE
are currently
being investigated from various points of view. 

An intriguing link to CFT has been examined
by Bauer and Bernard \cite{BB} (see also \cite{FW}) in which the SLE
differential equation is associated to a particular random walk on the 
Virasoro group. The relationship can be made more direct by establishing
a connection between the representation theory of CFT
and entities conserved in mean under the stochastic
process. This is based on the existence of level-two
singular vectors in highest-weight modules of the Virasoro algebra.

The approach of Bauer and Bernard has been extended from
ordinary CFT and SLE to $N=1$ superconformal field theory and stochastic
evolutions in $N=1$ superspace \cite{supersle,NR}, to logarithmic
CFT \cite{MRR,logsle}, and to CFT and SLE-type growth processes in 
smaller regions of the complex plane than ordinary chordal SLE \cite{LR}.
The present work offers
an extension from ordinary CFT to Wess-Zumino-Witten (WZW) models
where the conformal symmetry is supplemented by a Lie group symmetry.
We are thus led to consider stochastic evolutions of affine Lie
group transformations. Most of our results pertain to $SU(2)$
WZW models, and we refer to \cite{FMS} for a survey on CFT.

The present elevation to WZW models is discussed in the realm
of generating-function primary fields, see \cite{ZF,FGPP,PRY,thesis}.
They serve as a convenient way of handling the multiplet
of Virasoro primary fields comprising the $su(2)$ representation space
of a given conformal weight, as discussed below.
Transformations of the multiplet of Virasoro primary fields 
are replaced by transformations of the generating-function
primary field. As in ordinary CFT, these transformations may
be described in terms of group elements, here elements of the Virasoro group
and an affine extension of the Lie group $SU(2)$. Alternatively, they
may be described by utilizing that primary fields have tensor-like 
transformation properties, again as in ordinary CFT.
This allows one to link stochastic evolutions of the two kinds of group elements
to stochastic evolutions of conformal and affine 
transformations, respectively. These links may be established separately.
By combining them, one may relate the affine Sugawara construction
of Virasoro modes (see \cite{HKOC} and references therein)
to entities conserved in mean under the combined stochastic process.
This process is somewhat formal, though, as it is based on an infinite-dimensional
Brownian motion. As the Knizhnik-Zamolodchikov (KZ) equations
are generated by one of the conditions appearing in the affine
Sugawara construction, they too may be linked to stochastic evolutions.

A brief review of certain aspects of
$SU(2)$ WZW models and generating-function primary fields
is given in Section 2. The general link between $SU(2)$ WZW models
and stochastic evolutions is discussed in Section 3, while Section 4
concerns the special case corresponding to the affine Sugawara construction 
and the KZ equations. Section 5 contains some concluding remarks.

\section{On $SU(2)$ WZW models}

We shall discuss WZW models from an algebraic point of view, and are
therefore not concerned with their Lagrangian formulation.
The conformal symmetry is generated by the Virasoro modes
satisfying the algebra
\ben
 [L_n,L_m]=(n-m)L_{n+m}+\frac{c}{12}n(n^2-1)\delta_{n+m,0}
\label{vir}
\een
Transformations generated by the
affine $su(2)_k$ Lie algebra are here referred to as affine transformations.
The algebra, including the commutators with the Virasoro modes, reads
\bea
 \left[J_{+,n},J_{-,m}\right]&=&2J_{0,n+m}+kn\delta_{n+m,0} \nn
 \left[J_{0,n},J_{\pm,m}\right]&=&\pm J_{\pm,n+m} \nn
 \left[J_{0,n},J_{0,m}\right]&=&\frac{k}{2}n\delta_{n+m,0} \nn
 \left[L_n,J_{a,m}\right]&=&-mJ_{a,n+m}
\label{aff}
\eea
The level of this affine algebra is indicated by $k$, and 
we shall assume that it is a non-negative integer.
The non-vanishing entries of the Cartan-Killing form of $su(2)$ 
are given by  
\ben
 \kappa_{00}=\hf,\ \ \ \ \ \ \ \ \kappa_{+-}=\kappa_{-+}=1
\label{CK}
\een
and appear as coefficients to the central terms in (\ref{aff}).
Its inverse is given by
\ben
 \kappa^{00}=2,\ \ \ \ \ \ \ \ \kappa^{+-}=\kappa^{-+}=1
\label{CKinv}
\een
and comes into play when discussing the affine Sugawara construction below.

Virasoro primary fields are defined by their simple
transformation properties with respect to the Virasoro algebra:
\ben
 [L_n,\phi(z)]=\left(z^{n+1}\partial_z+\D(n+1)z^n\right)\phi(z)
\label{Lphi}
\een
Here $\D$ denotes the conformal weight of $\phi$.
The Virasoro primary fields of a given conformal weight $\D$ may be
organized in multiplets corresponding to spin-$j$ 
representations of the $su(2)$ algebra generated by $\{J_{a,0}\}$, where 
\ben
 \D=\frac{j(j+1)}{k+2}
\label{Dj}
\een
We shall label the members of such a multiplet as in
\ben
 \phi_{j,-j}(z),\ \phi_{j,-j+1}(z),\ ...,\ \phi_{j,j-1}(z),\ \phi_{j,j}(z)
\label{phij}
\een
The field $\phi_{j,m}$ has $J_{0,0}$ eigenvalue $m$, while a convenient
choice of relative normalizations of the fields is indicated by
\bea
 \left[J_{+,0},\phi_{j,m}(z)\right]&=&(j+m+1)\phi_{j,m+1}(z)\nn
 \left[J_{0,0},\phi_{j,m}(z)\right]&=&m\phi_{j,m}(z)\nn
 \left[J_{-,0},\phi_{j,m}(z)\right]&=&(j-m+1)\phi_{j,m-1}(z)
\label{Japhijm}
\eea

A generating function for these Virasoro primary fields may be written
\ben
 \phi(z,x)=\sum_{n=0}^{2j}x^n\phi_{j,j-n}(z)
\label{phizx}
\een
Since this field merely is a 
linear combination of Virasoro primary fields of the same conformal
weight, it too transforms as in (\ref{Lphi}). That is, the transformation
generated by the Virasoro group element $G$ simply reads
\ben
 G^{-1}\phi(z,x)G=(\partial_zf(z))^{\D}\phi(f(z),x)
\label{GphiG}
\een
for some conformal map $f$.

The action of the affine generators on the generating-function
primary field reads
\ben
 [J_{a,n},\phi(z,x)]=z^nD_a(x)\phi(z,x)
\label{Jphi}
\een
where the differential operators $D_a$ are defined by
\ben
 D_+(x)=-x^2\partial_x+2jx,\ \ \ \ \ \ D_0(x)=-x\partial_x+j,\ \ \ \ \ \ D_-(x)=\partial_x
\label{D}
\een
The set $\{-D_a\}$ generates the Lie algebra $su(2)$.
To derive the transformations generated by affine $SU(2)$
group elements, we first note that
\ben
 e^{-A}Be^A=e^{-ad_A}B
\label{eAB}
\een
where $ad_AB=[A,B]$. Using this, one finds that
\bea
 e^{-uJ_{+,n}}\phi(z,x)e^{uJ_{+,n}}&=& 
   (1-uz^nx)^{2j}\phi(z,\frac{x}{1-uz^nx})\nn
 e^{-uJ_{0,n}}\phi(z,x)e^{uJ_{0,n}}&=& 
   e^{-ujz^n}\phi(z,e^{uz^n}x)\nn
 e^{-uJ_{-,n}}\phi(z,x)e^{uJ_{-,n}}&=&\phi(z,x-uz^n)  
\label{eJ}
\eea
It follows that a general affine $SU(2)$ group element $U$
generates the transformation
\ben
 U^{-1}\phi(z,x)U=(\partial_xy(z,x))^{-j}\phi(z,y(z,x))
\label{UphiU}
\een
where $y(z,x)$ is a M\"obius transformation of $x$ with
$z$-dependent coefficients:
\ben
 y(z,x)=\frac{a(z)x+b(z)}{c(z)x+d(z)},\ \ \ \ \ \ \ \ a(z)d(z)-b(z)c(z)\neq0
\label{y}
\een

We shall also be interested in combinations of transformations generated
by Virasoro and affine $SU(2)$ group elements. In particular, a tranformation
generated by one type of group element followed by a transformation
generated by a group element of the other type results in
\ben
 G^{-1}U^{-1}\phi(z,x)UG=
   (\partial_zf(z))^{\D}(\partial_xy(z,x))^{-j}\phi(f(z),y(z,x))
\label{phiUG}
\een
or
\ben
 U^{-1}G^{-1}\phi(z,x)GU=
   (\partial_zf(z))^{\D}(\partial_xy(f(z),x))^{-j}\phi(f(z),y(f(z),x))
\label{phiGU}
\een
 
The affine Sugawara construction of the Virasoro modes in terms of
the affine generators is given by
\ben
 L_N=
   \frac{1}{2(k+2)}\kappa^{ab}\left(\sum_{n\leq-1}J_{a,n}J_{b,N-n}
  +\sum_{n\geq0}J_{a,N-n}J_{b,n}\right)
\label{sug}
\een
Here and in the following we shall use the convention of
summing over appropriately repeated group indices, $a=\pm,0$.
Acting on a highest-weight state, the affine Sugawara
construction gives rise to singular vectors of the combined
algebra. The decoupling of these is trivial for $N>0$ while for $N=0$
it merely reproduces the relation (\ref{Dj}). The condition
corresponding to $N=-1$ leads to the celebrated
KZ equations used in discussions of correlation functions.

\section{$SU(2)$ WZW models and stochastic evolutions}

By considering the Ito differential of both sides of
(\ref{GphiG}) where $G$ and $f(z)$ now are considered as stochastic
processes, $G_t$ and $f_t(z)$,
one may relate stochastic differentials of Virasoro group elements 
to stochastic evolutions of conformal maps.
We shall allow higher-dimensional Brownian motion satisfying
\ben
 dB_t^{\mu}dB_t^{\nu}=\delta^{\mu\nu}dt,\ \ \ \ \ \ \ \
   dB_t^{\mu}dt=dtdt=0
\label{dBdB}
\een
and $B_0^\mu=0$. We then have
\ben
 G_t^{-1}dG_t=\al_t(L)dt+\sum_{\mu}\beta_{\mu,t}(L)dB_t^{\mu},
   \ \ \ \ \ \ \ \ G_0=1
\label{dG}
\een
where $\al_t$ and $\beta_{\mu,t}$ are expressions in the Virasoro
modes. Similarly, the stochastic evolution of the associated
conformal maps may be written
\ben
 df_t(z)=f_{0,t}(z)dt+\sum_\mu f_{\mu,t}(z)dB_t^{\mu},
   \ \ \ \ \ \ \ \ \ f_0(z)=z
\label{df}
\een
Techniques for computing and comparing the Ito
differentials of both sides of (\ref{GphiG}) are
described in \cite{supersle,NR,logsle}.
With 
\bea
  \beta_{\mu,t}(L)&=&\sum_{n\in\Z}l_{\mu,n,t}L_n\nn
  \al_{0,t}(L)&=&\al_t(L)-\hf\sum_{\mu,\nu}\delta^{\mu\nu}
     \beta_{\mu,t}(L)\beta_{\nu,t}(L)
   \ =\ \sum_{n\in\Z}l_{0,n,t}L_n
  \label{al0}
\eea
one finds 
\bea
 f_{\mu,t}(z)&=&-\sum_{n\in\Z}l_{\mu,n,t}(f_t(z))^{n+1}\nn
 f_{0,t}(z)&=&-\sum_{n\in\Z}l_{0,n,t}(f_t(z))^{n+1}
   +\hf\sum_{\mu,\nu}\delta^{\mu\nu}\sum_{n,m\in\Z}
   (m+1)l_{\mu,n,t}l_{\nu,m,t}(f_t(z))^{n+m+1}
\label{fmu}
\eea
This provides a general link expressing the stochastic
evolution of conformal maps in terms of the stochastic 
Virasoro differentials.

A similar description of stochastic evolutions of the M\"obius transformations
(\ref{y}) in terms of affine $SU(2)$ differentials follows from an
evaluation of
the Ito differential of both sides of (\ref{UphiU}). For later convenience,
we shall base the analysis on a potentially non-diagonal
higher-dimensional Brownian motion with $W_t^\rho$ an invertible
linear combination of Brownian motions, satisfying
\ben
  dW_t^{\rho}dW_t^{\sigma}=\la^{\rho\sigma}dt,\ \ \ \ \ \ \ 
   dW_t^{\rho}dt=dtdt=0
\label{la}
\een
and $W_0^\rho=0$. Here $\la$ is a symmetric and invertible matrix. 
Using the same approach as above, we write
\ben
  U_t^{-1}dU_t=p_t(J)dt+\sum_{\rho}q_{\rho,t}(J)dW_t^{\rho},
   \ \ \ \ \ \ \ \ U_0=1
\label{dU}
\een
and 
\ben
 dy_t(z,x)=y_{0,t}(z,x)dt+\sum_\rho y_{\rho,t}(z,x)dW_t^{\rho},
   \ \ \ \ \ \ \ \ \ y_0(z,x)=x
\label{dy}
\een
The analogue to (\ref{al0}) reads
\bea
  q_{\rho,t}(J)&=&\sum_{n\in\Z}j_{\rho,n,t}^aJ_{a,n}\nn
  p_{0,t}(J)&=&p_t(J)-\hf\sum_{\rho,\sigma}\la^{\rho\sigma}
    q_{\rho,t}(J) q_{\sigma,t}(J)
   \ =\ \sum_{n\in\Z}j_{0,n,t}^aJ_{a,n}
   \label{p0}
\eea
and a goal is to express $y_{0,t}(z,x)$ and $y_{\rho,t}(z,x)$ in terms of
$j_{0,n,t}^a$, $j_{\rho,n,t}^a$ and $y_t(z,x)$.
We thereby find the following general link
\bea
 y_{\rho,t}(z,x)&=&\sum_{n\in\Z}z^n\left((y_t(z,x))^2j_{\rho,n,t}^++
   y_t(z,x)j_{\rho,n,t}^0-j_{\rho,n,t}^-\right)\nn
 y_{0,t}(z,x)&=&\sum_{n\in\Z}z^n\left((y_t(z,x))^2j_{0,n,t}^++
   y_t(z,x)j_{0,n,t}^0-j_{0,n,t}^-\right)\nn
 &+&\hf\sum_{\rho,\sigma}\la^{\rho\sigma}\sum_{n,m\in\Z}z^{n+m}
   \left((y_t(z,x))^2j_{\rho,n,t}^++
   y_t(z,x)j_{\rho,n,t}^0-j_{\rho,n,t}^-\right)\nn
  &&\hspace{4cm}\times\left(2y_t(z,x)j_{\sigma,m,t}^+
   +j_{\sigma,m,t}^0\right)
\label{yrho}
\eea

It turns out that the links (\ref{fmu}) and (\ref{yrho})
are unaltered if one considers the
Ito differential of both sides of (\ref{phiUG}) based on a
combination of the two group actions. This is a priori not obvious
since one in that case should allow that some of the Brownian motions 
$B_t^\mu$, appearing in (\ref{dG}) and (\ref{df}),
and $W_t^\rho$, appearing in (\ref{dU}) and (\ref{dy}), 
may be related. One would thus have to supplement
(\ref{dBdB}) and (\ref{la}) by
\ben
 dB_t^\mu dW_t^\rho=\hat{\la}^{\mu\rho}dt
\label{hatla}
\een
where $\hat{\la}$
could be non-vanishing. As already indicated, however, all terms
proportional to $\hat{\la}$ cancel and one is left with the
separate Virasoro and affine $SU(2)$ links.
The rationale for making this consistency check is that we shall
use the product $UG$ in our discussion of the
affine Sugawara construction in the following.

\section{Affine Sugawara construction}

Our next objective is to relate a set of stochastic evolutions
to the affine Sugawara conditions (\ref{sug}). To this end, we 
consider a general stochastic process $F_t$ with Ito differential
\ben
 dF_t=u_tdt+\sum_l v_{l,t}dB^l_t
\label{dF}
\een
where $B_0^l=0$.
For sufficiently well-behaved functions $u_t$ and $v_{l,t}$,
the time evolution of the expectation value of $F_t$ is given by
\ben
 \partial_tE[F_t]=E[u_t]
\label{E}
\een
and is seen to vanish provided $E[u_t]$ vanishes.
A goal is thus to find processes $F_t$ whose associated $u_t$'s
(\ref{dF}) correspond to the affine Sugawara conditions.
This illustrates a general scenario in which the
representation theory or structure of the CFT (here the $SU(2)$
WZW model) allows one
to put an entity equal to zero (here represented by (\ref{sug}) and $u_t$), 
thereby producing a martingale (here the stochastic
process $F_t$) of the system.

Since the affine Sugawara conditions involve both Virasoro and affine 
$su(2)$ generators, we ought to look for combinations of Virasoro
and affine $SU(2)$ group elements.
We therefore consider
\bea
 (U_tG_t)^{-1}d(U_tG_t)&=&G_t^{-1}(U_t^{-1}dU_t)G_t+G_t^{-1}dG_t+
   G_t^{-1}(U_t^{-1}dU_t)G_t(G_t^{-1}dG_t)\nn
  &=&\left(G_t^{-1}p_t(J)G_t+\al_t(L)+\sum_{\rho,\mu} \hat{\la}^{\rho\mu}
    G_t^{-1}q_{\rho,t}(J) G_t\beta_{\mu,t}(L)\right)dt\nn
   &+&\left(\sum_\mu\beta_{\mu,t}(L)\right) dB_t^\mu
    +\left(\sum_\rho G_t^{-1}q_{\rho,t}(J) G_t\right)dW_t^\rho
\label{dUG}
\eea
and
\bea
 (G_tU_t)^{-1}d(G_tU_t)&=& U_t^{-1}(G_t^{-1}dG_t)U_t+U_t^{-1}dU_t+
   U_t^{-1}(G_t^{-1}dG_t)U_t(U_t^{-1}dU_t)\nn
  &=&\left(U_t^{-1}\al_t(L) U_t+p_t(J)+
    \sum_{\mu,\rho}\hat{\la}^{\mu\rho}U_t^{-1}\beta_{\mu,t}(L)
   U_tq_{\rho,t}(J)\right)dt\nn
  &+&\left(\sum_\mu U_t^{-1}\beta_{\mu,t}(L) U_t\right)dB_t^\mu+
    \left(\sum_\rho q_{\rho,t}(J)\right) dW_t^\rho
\label{dGU}
\eea
It is noted that the inter-relating matrix $\hat{\la}$ (\ref{hatla}) appears in
these expressions.
Since the affine Sugawara conditions are linear in the Virasoro
modes and bilinear in the affine $su(2)$ modes, we shall 
work with the differential (\ref{dUG}) and not (\ref{dGU}).
The linearity in the Virasoro modes also suggests that we should consider
$\beta_{\mu,t}(L)=0$ and $\al_t(L)=\al_{0,t}(L)=L_N$, in which
case 
\ben
 G_t=e^{tL_N},\ \ \ \ \ \ \ \ G_t^{-1}dG_t=L_Ndt
\label{GLN}
\een
and
\ben
 \partial_t E[U_tG_t]=E[U_tG_t(L_N+G_t^{-1}p_t(J)G_t)]
\label{EUG}
\een
We should thus require that $p_{0,t}(J)=0$ and
\bea
 \sum_{\rho,\sigma}\la^{\rho\sigma}q_{\rho,t}(J)q_{\sigma,t}(J)
  &=& \frac{\kappa^{ab}}{k+2}\left(\sum_{n\leq-1}J_{a,n}J_{b,N-n}
  +\sum_{n\geq0}J_{a,N-n}J_{b,n}\right)\nn
 &=&\delta_{N,0}\frac{\kappa^{ab}}{k+2}
   \left(J_{a,0}J_{b,0}+2\sum_{n\geq1}J_{a,-n}J_{b,n}\right)\nn
 &+&(1-\delta_{N,0})\frac{\kappa^{ab}}{k+2}\sum_{n,m\in\Z}
   \delta_{n+m,N}J_{a,n}J_{b,m}
\label{qq}
\eea
The summation indices $\rho$ and $\sigma$
are then naturally considered as double indices:
$\rho=(a,n)$ where $a$ is a group index taking the values
$a=0,\pm$, and $n$ is an integer. That is,
\ben
 \sum_{\rho,\sigma}\la^{\rho\sigma}q_{\rho,t}(J)q_{\sigma,t}(J)
  =\sum_{n,m\in\Z}\la^{(a,n)(b,m)}
   q_{(a,n),t}(J)q_{(b,m),t}(J)
\label{rhoan}
\een
and the condition (\ref{qq}) for $N\neq0$ is satisfied if
\ben
 \la^{(a,n)(b,m)}=\kappa^{ab}\delta_{n+m,N},\ \ \ \ \ \ \ \ \ 
  q_{(a,n),t}(J)=\frac{1}{\sqrt{k+2}}J_{a,n}
\label{laq}
\een
The $N=0$ condition is not covered by this analysis due
to the divergencies appearing in the affine Sugawara construction
when the normal ordering of the affine modes (as in 
(\ref{sug}) and (\ref{qq})) is omitted. Since the Ito calculus is
based on a {\em symmetric} 'two-form' 
in (\ref{la}), $\la^{\rho\sigma}=\la^{\sigma\rho}$, 
we are confined to ordinary products
and thus seem deprived of the power of normal ordering required
in the $N=0$ condition.

The stochastic differential equation of the affine $SU(2)$ group element 
accompanying (\ref{GLN}) for $N\neq0$ is now seen to be
\ben
 U_t^{-1}dU_t=\frac{\kappa^{ab}}{2(k+2)}
     \sum_{n\in\Z}J_{a,N-n}J_{b,n}dt
  +\frac{1}{\sqrt{k+2}}\sum_{n\in\Z}J_{a,n}dW_t^{(a,n)},\hspace{1cm}
   U_0=1
\label{UUB}
\een
This is a somewhat formal expression as it involves an
infinite-dimensional Brownian motion.
It is emphasized that there is a pair $(G_t,U_t)$ for each of 
the affine Sugawara conditions (\ref{sug}) with $N\neq0$.
An explicit indication of which condition such a pair refers
to has nevertheless been omitted for notational reasons.

The conformal maps $f_t(z)$ associated to the Virasoro group elements
$G_t$ given in (\ref{GLN}) evolve deterministically as we have
\ben
 l_{\mu,n,t}=0,\ \ \ \ \ \ \ l_{0,n,t}=\delta_{n,N}
\label{ll}
\een
and subsequently from (\ref{fmu})
\ben
 df_t(z)=-(f_t(z))^{N+1}dt,\ \ \ \ \ \ \ f_0(z)=z
\label{dfdet}
\een
This is solved by
\ben
 f_t(z)=\frac{z}{(1+Ntz^N)^{1/N}},\ \ \ \ \ \ \ \ \ N\neq0
\label{fdet}
\een
or
\ben
 f_t(z)=  ze^{-t}, \ \ \ \ \ \ \ \ \ \ \ N=0 
\label{fdet0}
\een
and can be verified directly using (\ref{Lphi}), (\ref{GphiG}) and (\ref{eAB}). 

To determine the stochastic differentials of the M\"obius
transformations $y_t(z,x)$ associated to (\ref{UUB}), we
rely on the link (\ref{yrho}). We have
\ben
 j_{(a,n),m,t}^b=\frac{1}{\sqrt{k+2}}\delta_{nm}\delta_a^b,\ \ \ \ \ \ \ \ 
 j_{0,n,t}^a=0
\label{jj}
\een
and it follows from (\ref{yrho}) that
\ben
 y_{(a,n),t}=\frac{1}{\sqrt{k+2}}z^n
   \left(y^2\delta_a^++y\delta_a^0-\delta_a^-\right),\ \ \ \ \ \ \ 
   y_{0,t}=0
\label{yy}
\een
and hence
\ben
 dy_t(z,x)=\frac{1}{\sqrt{k+2}}\sum_{n\in\Z}z^n
  \left(y^2dW_t^{(+,n)}+ydW_t^{(0,n)}-dW_t^{(-,n)}\right),\ \ \ \ \ \ \ y_0(z,x)=x
\label{dysto}
\een
Note that the dependence on $N$ is given implicitly
via $dW_t^{(a,n)}dW_t^{(b,m)}=\kappa^{ab}\delta_{n+m,N}dt$, cf.
(\ref{laq}).

In order to express $dy_t(z,x)$ in terms of ordinary, though infinitely many, 
'orthonormal' Brownian motions instead of $\{W^{(a,n)}\}$, 
we perform some linear transformations.
First we introduce the linear combinations
\bea
 B_t^{(+,n)}&=&\frac{1}{\sqrt{2}}\left(B_t^{(1,n)}+iB_t^{(2,n)}\right)\nn
 B_t^{(0,n)}&=&\sqrt{2}B_t^{(3,n)}\nn
 B_t^{(-,n)}&=&\frac{1}{\sqrt{2}}\left(B_t^{(1,n)}-iB_t^{(2,n)}\right)
\label{Ba}
\eea
where we have introduced an infinite set of Brownian motions
labeled as $B_t^{(\ell,n)}$. They satisfy the orthonormality conditions
\ben
 dB_t^{(\ell,n)}dB_t^{(\ell',n')}=\delta_{\ell\ell'}\delta_{nn'}dt,\ \ \ \ \ \ \ \ \ \ 
  \ell,\ell'\in\{1,2,3\},\ \ \ n,n'\in\Z
\label{ell}
\een
and $B_0^{(\ell,n)}=0$.
With our standard convention for the group index, $a=\pm,0$,
we then have
\bea
   W_t^{(a,n>N/2)}&=&
   \frac{1}{\sqrt{2}}\left(B_t^{(a,n)}+iB^{(a,N-n)}\right)\nn
  W_t^{(a,n=N/2)}&=&
      B^{(a,N/2)},\hspace{3cm} {\rm for}\ N\ {\rm even}\nn
  W_t^{(a,n<N/2)}&=&
     \frac{-i}{\sqrt{2}}\left(B_t^{(a,n)}+iB^{(a,N-n)}\right)
\label{N}
\eea
These processes are seen to respect 
$dW_t^{(a,n)}dW_t^{(b,m)}=\la^{(a,n)(b,m)}dt$ with $\la$ given in
(\ref{laq}). It is now straightforward to express $dy_t(z,x)$ given in
(\ref{dysto}) in terms of the orthonormal Brownian 
differentials $dB_t^{(\ell,n)}$, and we find
\bea
 dy_t(z,x)&=&\frac{1}{2\sqrt{k+2}}\sum_{n>N/2}
   (z^n+z^{N-n})
   \left((y^2-1)dB_t^{(1,n)}+i(y^2+1)dB_t^{(2,n)}
     +2ydB_t^{(3,n)}\right)\nn
   &+&\frac{i}{2\sqrt{k+2}}\sum_{n<N/2}(z^{N-n}-z^{n})
  \left((y^2-1)dB_t^{(1,n)}+i(y^2+1)dB_t^{(2,n)}
     +2ydB_t^{(3,n)}\right)\nn
  &+&\frac{(1+(-1)^N)}{2\sqrt{2(k+2)}}z^{N/2}
   \left((y^2-1)dB_t^{(1,n)}+i(y^2+1)dB_t^{(2,n)}
     +2ydB_t^{(3,n)}\right)\nn
 y_0(z,x)&=&x
\label{dyB}
\eea
where the last term proportional to $z^{N/2}$
vanishes for $N$ odd. It is recalled that
$N\neq0$ in this expression. The condition underlying the KZ
equations is easily obtained by setting $N=-1$.

\section{Conclusion}

We have discussed how $SU(2)$ WZW models may be linked to
stochastic evolutions. The conformal symmetry is thereby related
to stochastic evolutions of conformal maps whereas the affine
$SU(2)$ invariance is linked to stochastic evolutions of affine transformations.
This extends work done in \cite{BB,LR,supersle,NR,MRR,logsle}
on connections between CFT and SLE, and generalizations thereof.
An objective of the present work was to develop a set of stochastic differential equations
corresponding to the affine Sugawara construction of the Virasoro generators
in terms of the affine generators. This has been achieved for all
Virasoro modes but $L_0$, in which case our approach seems to
be incapable of reproducing the required normal ordering of the affine modes.
Since the $L_{-1}$ mode is covered, we have thus obtained
a stochastic differential equation describing the condition underlying
the KZ equations. The associated stochastic process is somewhat formal, as it
is based on an infinite-dimensional Brownian motion. We nevertheless hope
that our analysis may prove itself useful.
\vskip.5cm
\noindent{\em Acknowledgements}
\vskip.1cm
\noindent The author is grateful to 
P. Jacob for discussions at early stages of this work, and thanks
M.B. Halpern for comments.

\end{document}